# Calibration of hydrogen atoms measurement using femtosecond two-photon laser induced fluorescence

Andrey Starikovskiy[1], Arthur Dogariu[1,2]

[1]Dept. of Mechanical and Aerospace Engineering, Princeton University
[2]Dept. of Aerospace Engineering, Texas A&M University

*Abstract*

A new calibration method for H-fs-TALIF is proposed, and the ratio of two-photon absorption cross-sections, $\sigma^{(2)}$, for atomic hydrogen (H) and krypton (Kr) is determined using the broadband emission of a femtosecond laser system. The estimated ratio of the two-photon absorption cross-sections for H and Kr is $\sigma^{(2)}(Kr)/\sigma^{(2)}(H) = 0.027 \pm 20\%$, which is nearly twenty times lower than values previously obtained with narrowband nanosecond lasers. This discrepancy highlights the need for independent calibration of TALIF measurements for a specific laser pulsewidth range.

*Introduction*

TALIF extends the applicability of laser-induced fluorescence (LIF) to light atoms (such as O, N, and H) with wavelength transitions in the VUV range by employing a two-photon excitation scheme. The advantages of femtosecond TALIF (fs-TALIF) over the traditional nanosecond TALIF technique have been recently reported [1,2]. These advantages include greater excitation efficiency, higher fidelity quenching rate measurements at red-shifted UV wavelengths over a wider range of operating conditions, significantly reduced photolytic interference, no quenching during excitation, and a much higher measurement rate (in the kHz range). The use of femtosecond (fs) and picosecond (ps) lasers is particularly beneficial for TALIF when measuring the effective lifetime of excited states in highly collisional environments. Fs lasers allow excitation on a time scale much shorter than the de-excitation time by quenching, enabling a clear separation of the population process of radiating states from their depopulation (via collisions and radiation).

This, in turn, facilitates the measurement of the ratios of different quenching mechanisms during experiments.

Figure 1 illustrates the simplified energy diagram of the TALIF processes for atomic hydrogen and krypton. For atomic hydrogen, a two-photon transition occurs from the ground state H(1s-$^2S_{1/2}$) to the excited state H(3d-$^2D_{3/2,5/2}$) at a wavelength of $\lambda$ = 205.08 nm when exposed to intense coherent radiation. This excited state is depopulated by both collisional and radiative mechanisms. Collisional quenching is highly dependent on the composition and density of the medium where the measurements are conducted. Radiative depopulation results in a transition to the H(2p-$^2P_{1/2,3/2}$) state, emitting a photon at a wavelength of 656.3 nm.

For krypton, two-photon excitation from the ground state Kr(4$p^6$-$^1S_0$) occurs at a wavelength of $\lambda$ = 204.13 nm, transitioning the atom to the excited state Kr(5$p'$[3/2]$_2$). This is followed by a radiative transition to the Kr(5$s'$[1/2]$_1$) state, with photon emission at a wavelength of 826.3 nm.

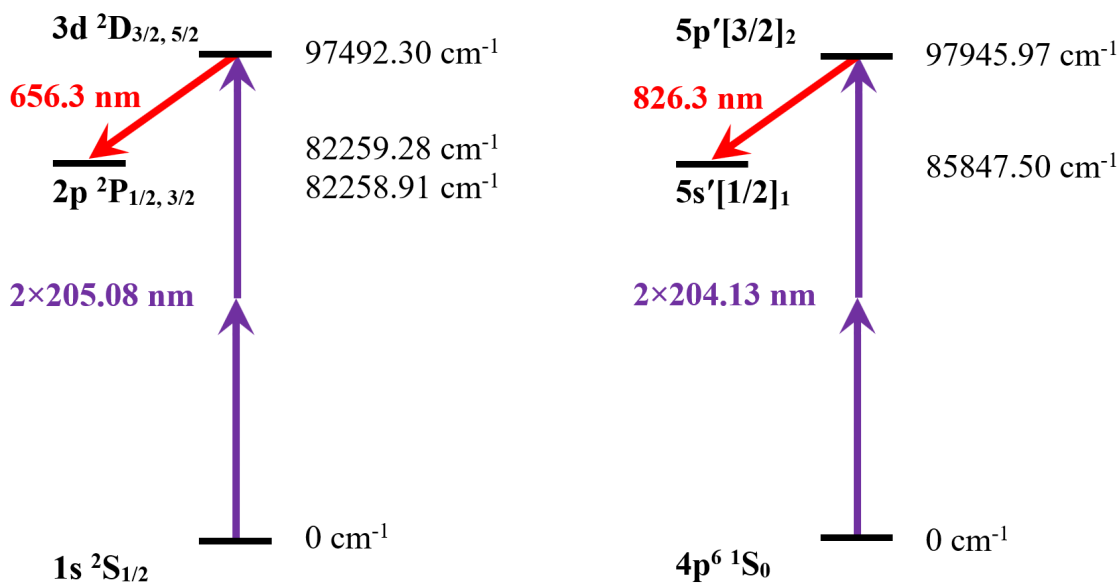

Figure 1. Two-photon excitation energy diagram for TALIF transitions in hydrogen (left) and krypton (right).

Since the two-photon excitation wavelengths for H and Kr atoms are close, tuning the laser to a different frequency does not significantly affect beam geometry or focusing. This greatly simplifies the absolute calibration of hydrogen atomic concentration measurements using the known krypton atomic concentration in the calibration experiment.

Figure 1 presents a simplified structure of hydrogen and krypton levels. For example, $3s^2S_{1/2}$, $3d^2D_{3/2}$ and $3d^2D_{5/2}$ levels of hydrogen are within less than 0.0004 nm (with respect to the two-photon excitation wavelength), so it is inevitable that 3s and 3d states are excited simultaneously. 3s states can be ignored in the context of H-TALIF measurements because the transition probability for 2p-3s is an order of magnitude smaller than the transition probability for 2p-3d [3].

The diagram for transitions in krypton (Figure 1, right), states that Kr is excited to the $4s^24p^5(^2P°_{1/2})5p'\ ^2[^3/_2]_2$ level only. It should be noted, however, that in the vicinity of this level there is the $4s^24p^5(^2P°_{1/2})5p'\ ^2[^1/_2]_1$ level. But transition to $5s'\ ^2[^1/_2]_1 - 5p'\ ^2[^1/_2]_1$ is 2 nm away from the transition to $5s'\ ^2[^1/_2]°_1 - 5p'\ ^2[^3/_2]_2$ and it has two times smaller transition probability [4]. That is why the excitation of the $5p'\ ^2[^1/_2]_1$ level can be ignored in the context of H-TALIF measurements, too.

For sufficiently long excitation pulses (in nanosecond range) with the results of the relative measurements of the TALIF signals in pure Kr and the gas containing H atoms, the density of the H neutrals, $n(H)$, can be calculated from the measured signals $S(H)$ for hydrogen atoms (at $\lambda$ = 656.3 nm) and $S(Kr)$ for krypton atoms (at $\lambda$ = 826.3 nm) by knowing the density of the reference gas, $n(Kr)$, as follows [5]:

$$n(H) = n(Kr) \frac{T(Kr)}{T(H)} \frac{\eta(Kr)}{\eta(H)} \frac{\sigma^{(2)}(Kr)}{\sigma^{(2)}(H)} \frac{a_{2\to3}(Kr)}{a_{2\to3}(H)} \frac{S(H)}{S(Kr)} kt \qquad (1)$$

Here, $T$ and $\eta$ represent the wavelength-dependent transmission of the optical filters used and the detector efficiency at 656.3 nm and 826.3 nm, respectively. $\sigma^{(2)}$ denotes the two-photon absorption cross-sections at 205.08 nm for hydrogen and 204.13 nm for krypton. $a_{2\to3}$ is a branching ratio:

$$a_{2\to3} = A_{2\to3}/(A_2 + Q_2) \qquad (2)$$

where $A_{2\to3}$ is an Einstein coefficient, $A_2$ is the inverse radiative lifetime $\tau$ ($A_2 = 1/\tau$), and $Q_2$ is the inverse quenching lifetime $Q_2 = k_q \times N$, where $k_q$ is the collisional quenching rate. The term $kt$ was added to account for the gating time ($t_2 - t_1$) and delay $t_1$ (relative to the laser pulse) of the ICCD camera. This term can be calculated by integrating the laser signal $I(t)$ at a given wavelength over the gating time with the appropriate delay time:

$$kt = \frac{\int_{t_1}^{t_2} I^2(t)dt \text{ (Kr)}}{\int_{t_1}^{t_2} I^2(t)dt \text{ (H)}} \tag{3}$$

It should be noted, however, that this simplified description of the atomic response using the process cross-section is an approximation, applicable only in limiting cases where a stationary regime is reached. This is generally valid on longer time scales, typically exceeding 1 ns (see, e.g., [6]). For pulses shorter than 90 ps, the observed fluorescence is influenced by the onset of Rabi oscillations from the ground to the excited level [7].

The use of ultrashort pulses leads to an increase in the cross-section ratio at resonance, with the ratio of the emission lifetimes of the two atoms acting as the parameter. Since the excited Kr atom has a longer lifetime than the excited H atom, it follows that the use of ultrashort pulses should result in a reduction in the cross-section ratio.

It is important to note that the simple scaling laws described above are only valid in the low-intensity regime, at the onset of Rabi oscillations. In the high-intensity regime, the population of the excited level no longer increases monotonically with the excitation time. When the laser pulse duration is shortened and the excitation remains sufficient to populate the excited state, the secular approximation becomes invalid, and the optical Bloch equations no longer accurately represent the dynamics of the excited state population. The dynamics of excited state populations under ultrashort pulse excitation, including two-photon processes, have been extensively studied (see, for example, [8, 9]). In [10], it was observed that the signal was approximately 100 times stronger when excitation occurred on a nanosecond scale (around 3 mJ per pulse with a duration of 10 nanoseconds) compared to excitation on a femtosecond scale (about 6 μJ per pulse with a duration of 90 femtoseconds), despite the pulses having similar $E^2/t$ ratios.

However, since the purpose of this paper is to provide an experimental solution for measuring the absorption cross-section ratios, a simple approximation that enables TALIF measurements in the femtosecond regime will suffice, and we can apply the same approach used for steady-state conditions. For diagnostic purposes, equation (1) with cross-section ratio can still be useful, with the primary consideration being the precise value of this parameter.

It should be mentioned that the overlap between the laser emission profile and the absorption profile of a specific line is crucial, and this overlap varies significantly depending on line broadening and laser characteristics. Femtosecond lasers have very broad lines, allowing all atoms to participate in absorption. In contrast, the narrow-band emission of seeded nanosecond lasers excites only particles near the line center. While the total transition probability remains constant, the interaction between the atom and the laser is strongly influenced by line broadening. Based on the fact that the femtosecond laser bandwidth greatly exceeds any broadened atomic transition, we expect short pulse laser TALIF to be much less sensitive to experimental conditions such temperature, pressure, mixture composition than the nanosecond TALIF experiments.

Thus, the ratio $\sigma^{(2)}(Kr)/\sigma^{(2)}(H)$ for femtosecond and nanosecond laser systems can differ due to variations in laser spectral line widths and the differing overlap of emission profiles with atomic line profiles. The Doppler broadening of atomic hydrogen, which has a mass 84 times smaller than krypton, significantly increases the radiation absorption cross-section for the broadband femtosecond system compared to the nanosecond system. Note also that, for pairs such as O-Xe, transitioning from narrowband nanosecond to broadband femtosecond excitation can lead to calibration corrections due to different broadening and distinct excitation dynamics in the femtosecond regime.

In particular, estimating atomic hydrogen concentration using approximate relations (1)-(3) relies on measuring the ratio of two-photon absorption cross-sections, $\sigma^{(2)}$, for the H and Kr pair. These measurements have been conducted several times using narrow-band nanosecond lasers [5,11]. In these experiments, the concentration of atomic hydrogen was determined by adding $NO_2$ and recording the OH* chemiluminescence from the reaction of H and $NO_2$. This approach introduced an uncertainty of ±50% in the estimated value of the two-photon excitation cross-section ratio [5]. The value of $\sigma^{(2)}(Kr)/\sigma^{(2)}(H)$ obtained in [5,11] using two-photon excitation with a narrow-band nanosecond laser was subsequently used for absolute calibration of hydrogen atom measurements with both narrow-band nanosecond lasers and broadband picosecond and femtosecond lasers [12-18].

It is therefore essential to develop a mechanism for experimental calibration of measurements to enable the use of short and ultrashort laser pulses for quantitative TALIF. This paper aims to introduce a new H-atom source for calibration and to emphasize that all absolute measurements require careful calibration.

*H-atoms source*

The most challenging aspect of achieving absolute calibration for H-TALIF is generating a known concentration of atomic hydrogen. In [5,11], this was addressed by adding $NO_2$ to the gas stream and analyzing the resulting fluorescence. However, this method is labor-intensive and has notable accuracy limitations [5]. In this work, we propose using a low-pressure DC discharge in xenon with a small addition of molecular hydrogen to generate a controlled amount of H atoms. The main concept behind using this discharge as a tool for H-TALIF calibration is the complete dissociation of molecular hydrogen in the xenon plasma. In such a system, the energy structure of xenon and atomic hydrogen levels prevents the accumulation of hydrogen atoms in excited states and hydrogen ionization. A detailed analysis of the kinetics of energy exchange and charge transfer mechanisms in the $H_2$-H-Xe mixture is presented below.

The discharge chamber consists of a long quartz tube surrounded by a grounded aluminum shield (Figure 2). The inner diameter of the discharge tube is $D = 50$ mm, and the length of the discharge gap between electrodes is $L = 1000$ mm. The electrode system is designed symmetrically, with both electrodes made of an aluminum tube of diameter $d = 30$ mm and length $l = 250$ mm. The working area of the electrodes is $S = 470$ cm$^2$. This geometry allows the use of a power supply of any polarity and alternating voltage without reducing plasma generation efficiency at low gas pressures within the discharge gap. The large electrode area enables operation over a wide range of discharge currents without transitioning to an anomalous discharge regime, which could decrease discharge homogeneity and local gas temperature increase near the electrodes.

At a typical voltage $U \sim 600$ V across the discharge cell and a total gas pressure $P \sim 10$ Pa, the reduced electric field in the discharge is $E/n \approx 200$ Td, ensuring efficient excitation and ionization of the gas in the discharge. The electrodes are separated from the aluminum shield by dielectric inserts that allow high voltages of up to 50 kV to be applied to the system. Direct measurements of voltage and discharge current in the cell were conducted using high-voltage, high-frequency probes. The high-voltage power supply, TREK-40, has a high internal impedance ($Z = 200$ kOhm) and delivers square-wave alternating current with a frequency of 1 kHz to the discharge cell.

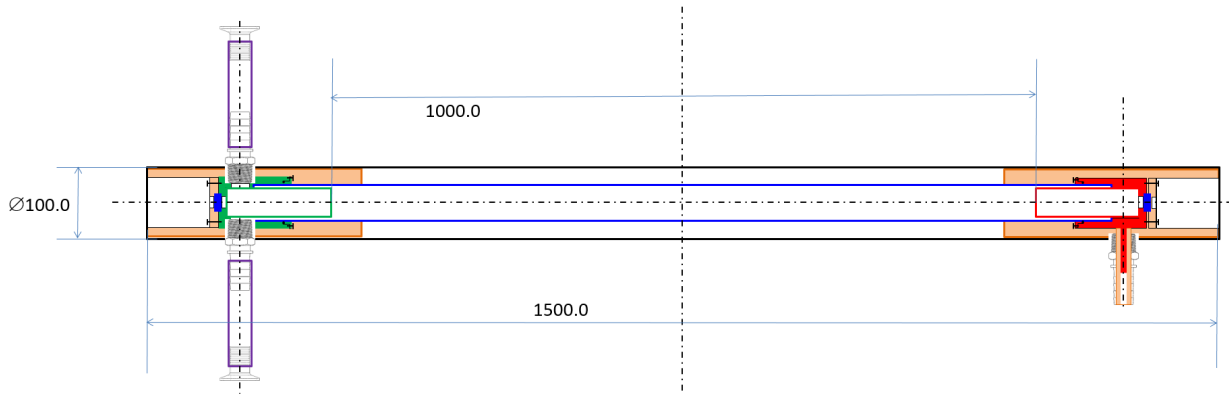

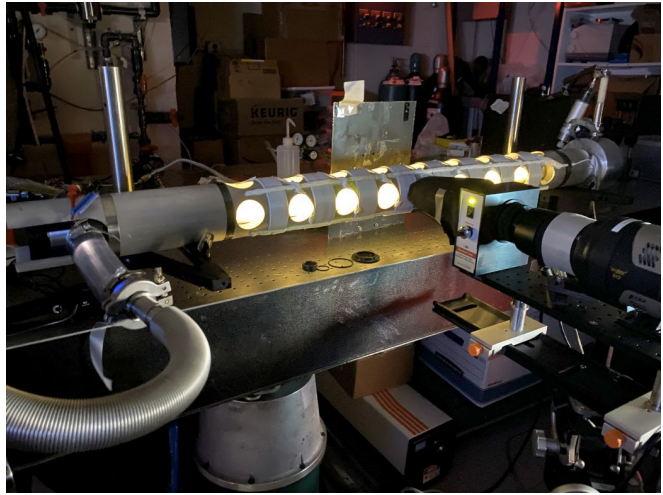

Figure 2. Discharge cell geometry and discharge development in the mixture
Xe + 112 ppm H$_2$. $P$ = 10 Pa.

The electrodes have quartz end windows with a diameter of 25 mm. The windows for optical diagnostics of the discharge are positioned in an aluminum shield at intervals of 150 mm (Figure 2). Connections for the vacuum system are mounted on the low-voltage electrode. The inlet and outlet nozzles are installed asymmetrically to provide continuous gas exchange in the discharge gap during the experiment. This is achieved by forming a near-wall flow of fresh gas directed into the discharge tube, while continuous gas pumping out occurs near the axis of the discharge cell. The vacuum system, equipped with a turbo pump, enables an initial vacuum of $P_0 \sim 10^{-5}$ Pa. Gas circulation within the discharge tube allows complete gas exchange every 2 seconds. Together with the vacuum system's leakage rate of approximately $\sim 2\times10^{-2}$ Pa/s, this setup ensures the mixture inside the discharge tube remains pure, at a level of impurities less than 5 ppm, under a total pressure of $P$ = 10 Pa.

*Measurements of the discharge parameters*

Figure 3 shows the results of current and voltage dynamics measurements in the discharge gap for mixtures with 112 and 200 ppm of $H_2$ in Xe. It can be observed that adding a small amount of hydrogen does not significantly affect the discharge parameters. The relatively slow recombination of the plasma preserves substantial ionization as the voltage passes through zero value. As a result, the current follows the voltage without delay, reaching self-consistent values that correspond to the instantaneous voltage in the gap within a time much shorter than the characteristic pulse length (Figure 3).

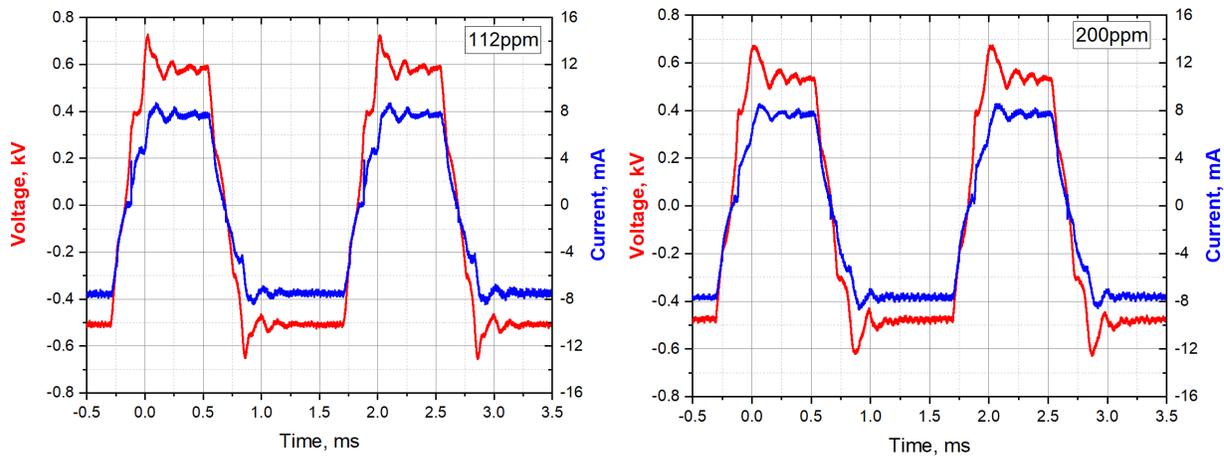

Figure 3. Measured dynamics of voltage and current in the discharge gap. Xe+$H_2$, $P = 10$ Pa. Left: mixture Xe + 112 ppm $H_2$; right: mixture Xe + 200 ppm $H_2$

Figure 4 shows the discharge power dynamics. The average discharge power in the positive half-wave is about $P \sim 4.5$ W and shows good cycle-by-cycle reproducibility. The power in the negative half-wave also shows the same stability and the average power is about $P \sim 3.8$ W. The small difference of the discharge power is attributed to the asymmetry of the initial high-voltage waveform (Figure 3). The effective impedance of the discharge also shows great stability and values $R \sim 75$ kOhm for both investigated mixtures (Figure 4). The relatively low average discharge power keeps the temperature of the discharge tube close to room conditions due to the intensive external air cooling. A more accurate estimate of the temperature increase is provided below.

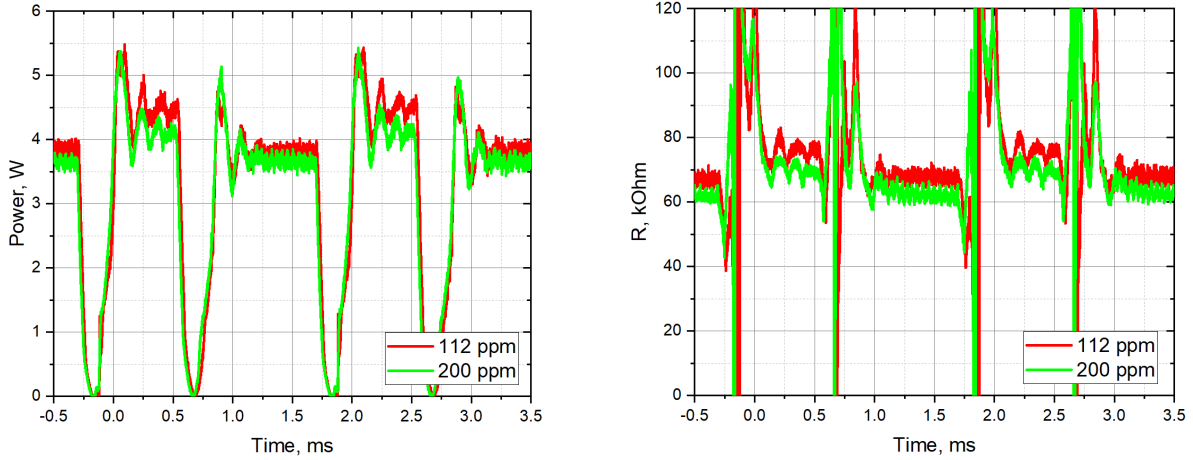

Figure 4. Left: Instantaneous discharge power for different mixtures. Right: Effective impedance of the discharge vs time. Xe+H$_2$, $P$ = 10 Pa.

*Plasmachemical processes in xenon-hydrogen mixtures*

The Xe-H$_2$ system possesses unique properties that allow for both the complete dissociation of molecular hydrogen in the mixture and the prevention of significant excitation and ionization of hydrogen atoms.

One of these unique properties is the hierarchy of ionization energies of the components (Table 1). Molecular hydrogen has a high ionization energy $E_i$ (H$_2$) = 15.426 eV, atomic hydrogen has an ionization energy of 13.599 eV, and xenon atom has the lowest ionization energy $E_i$ (Xe) = 12.13 eV (Table 1). This hierarchy ensures that no hydrogen ions remain in the system due to rapid charge transfer reactions—resulting in all ions in the system quickly becoming Xe$^+$ ions (reactions (4)-(6)).

$$\text{Xe} + e \rightarrow 2e + \text{Xe}^+ \tag{4}$$

$$\text{H}_2^+ + \text{Xe} \rightarrow \text{H}_2 + \text{Xe}^+ \tag{5}$$

$$\text{H}^+ + \text{Xe} \rightarrow \text{H} + \text{Xe}^+ \tag{6}$$

Electron impact ionization (reaction 4) is a very fast process in strong electric fields. The rate constant depends on the magnitude of the reduced electric field $k = f(E/n)$ and, under the conditions of this study, can be calculated by solving the Boltzmann equation in the local field approximation. For typical conditions of $E/n \sim$ 200 Td, the rate constant for electron-impact ionization of xenon is $k = 3\times 10^{-10}$ cm$^3$/s (see also the discussion below).

Table 1. Energy of electronically excited states of H and Xe atoms, ionization, and dissociation of molecular hydrogen [19].

| Component | State | Energy, cm$^{-1}$ | Energy, eV |
|---|---|---|---|
| Hydrogen molecule | Dissociation threshold | 36117.215 | 4.478 |
|  | Ionization threshold | 124418.8856 | 15.426 |
| Hydrogen atom | $2p(^2P^o_{1/2})$ | 82258.921 | 10.199 |
|  | $2s(^2S_{1/2})$ | 82258.956 | 10.199 |
|  | $2p(^2P^o_{3/2})$ | 82259.286 | 10.199 |
|  | $3p(^2P^o_{1/2})$ | 97492.213 | 12.088 |
|  | $3s(^2S_{1/2})$ | 97492.222 | 12.088 |
|  | $3p(^2P^o_{3/2})$, $3d(^2D_{3/2})$ | 97492.321 | 12.088 |
|  | $3d(^2D_{5/2})$ | 97492.357 | 12.088 |
|  | Ionization threshold | 109678.774 | 13.598 |
| Xenon atom | $6s[3/2]^o_2$ | 67068.047 | 8.315 |
|  | $6s[3/2]_1$ | 68045.663 | 8.437 |
|  | $5p^5(^2P^o_{1/2})$ $6s'[1/2]^o_0$ | 76197.292 | 9.447 |
|  | $6s'[1/2]^o_1$ | 77185.56 | 9.570 |
|  | $5p^5(^2P^o_{3/2})$, $6p[1/2]_1$ | 77269.649 | 9.580 |
|  | $6p[5/2]_2$ | 78120.303 | 9.686 |
|  | $6p[5/2]_3$ | 78403.562 | 9.721 |
|  | $6p[3/2]_1$ | 78956.538 | 9.789 |
|  | $6p[3/2]_2$ | 79212.97 | 9.821 |
|  | $6p[1/2]_0$ | 80118.962 | 9.933 |
|  | $5d[3/2]_1$ | 83889.971 | 10.401 |
|  | $7p[5/2]_2$ | 88469.213 | 10.969 |
|  | Ionization threshold | 97834.834 | 12.13 |

The charge transfer reaction rates are of the order of gas kinetic rates due to the interaction of the ion and the induced dipole. The rate constant for the non-resonant charge-transfer reactions (reactions 5 and 6) can be estimated using the "capture" model [20]:

$$k = 3\times 10^9 \, \pi \, e \, (\alpha/\mu)^{1/2} \, [\text{cm}^3/\text{s}],$$

where $\alpha$ is the polarizability of the xenon atom in [Å$^3$], $\alpha = 4.04$; $\mu$ is the reduced collision mass in atomic mass units (a.m.u.), and $e$ is the elementary charge in Coulombs [C].

The rate constant for the charge-transfer reaction involving the molecular hydrogen ion is estimated as $k_5 = 2\times 10^{-9}$ [cm$^3$/s], while for the atomic ion – due to the smaller reduced mass – it is $k_6 = 3\times 10^{-9}$ [cm$^3$/s] (according to the "capture" model [20]). These high rates of charge-transfer reactions indicate that there is no accumulation of molecular or atomic hydrogen ions in the system.

Another key feature of the Xe-H pair is the structure of the electronically excited levels of the hydrogen atom and xenon (Table 1). The lower excited states of the hydrogen atom have higher energy than those of xenon, making the collisional depopulation of the excited states of atomic hydrogen highly efficient:

$$\text{H} + e \rightarrow \text{H}^* + e \tag{7}$$

$$\text{H}^* + \text{Xe} \rightarrow \text{H} + \text{Xe}^* \tag{8}$$

The rate constants of collisional transfer of electron excitation in processes of type (8) have been measured multiple times. In [21] for the collisional quenching of H($n=3$) states in a xenon atmosphere the value $k_Q = 31\pm 2\times 10^{-10}$ cm$^3$/s was observed. In [5], for the H(3d $^2$D$_j$) state, an estimate of $k_Q = 19.8\times 10^{-10}$ cm$^3$/s was obtained. Thus, the competition between the population of electronically excited states of atomic hydrogen by electron impact (7) and the collisional depopulation of these states (8) in collisions with xenon atoms – considering the low concentration of atomic hydrogen in the gas (maximum 400 ppm) and the low degree of gas ionization in the discharge under these conditions – results in the rapid quenching of all electronically excited states of atomic hydrogen.

The excited electronic levels of xenon have energies above 8 eV, which leads to rapid dissociative collisional quenching of these states upon collision with molecular hydrogen, whose dissociation energy is less than 4.5 eV (Table 1).

$$Xe + e \rightarrow Xe^* + e \tag{9}$$

$$Xe^* + H_2 \rightarrow Xe + H + H \tag{10}$$

The rate of electronic state excitation by electron impact in plasma, as well as the ionization rate, strongly depends on the value of the reduced electric field $k = f(E/n)$ and, under the conditions of this study, can be calculated by solving the Boltzmann equation in the local field approximation. For typical conditions of $E/n \sim 200$ Td, the rate constant of excitation of the electronic states of xenon is nearly an order of magnitude higher than its ionization rate, with $k = 2 \times 10^{-9}$ cm$^3$/s (see below for a detailed discussion).

The upper excited states of xenon can be quenched by collisions with other xenon atoms, transferring part of the energy to lower electronic states. However, due to the large energy defect, the Xe(6s [3/2]$_2$) (8.315 eV) and Xe(6s [3/2]$_1$) (8.437 eV) levels are practically unaffected by collisions with xenon atoms.

The primary channel for depopulating these states in the gas phase becomes collisions with molecular hydrogen (reaction 10). The quenching rates of individual electronic states of xenon in collisions with hydrogen molecules are summarized in Table 2.

Table 2. Quenching rate coefficients for electronically-exited states of xenon in collisions with molecular hydrogen.

| Xe* state | Quenching rate | Reference |
| --- | --- | --- |
| 7p [5/2]$_2$ | $(4.5\pm0.5)\times10^{-10}$ cm$^3$/s | [22] |
| 5d [3/2]$_1$ | $(2.6\pm0.2)\times10^{-10}$ cm$^3$/s | [23] |
| 6p [1/2]$_0$ | $(6.0\pm0.8)\times10^{-10}$ cm$^3$/s | [24] |
| 6s [3/2]$_1$ | $(0.86\pm0.04)\times10^{-10}$ cm$^3$/s | [25] |

Note that the lowest quenching rate of the electronically excited state of xenon by molecular hydrogen is nearly $10^{-10}$ cm$^3$/s. The competition between molecular hydrogen dissociation in such collisions, three-body volumetric recombination of H atoms, and their recombination on the wall will be discussed below.

In a mixture of xenon with a small addition of hydrogen under low pressure conditions, where three-body processes can be neglected, the primary ion is Xe$^+$. Electron-impact excitation of xenon electronic states introduces an effective dissociation channel for molecular hydrogen. The excitation of atomic hydrogen electronic states in the discharge via direct electron impact is rapidly quenched through collisions with xenon atoms, which have lower excitation energies of electronic states.

At low hydrogen concentrations and low mixture pressures, the reaction of gas-phase three-body recombination

$$H + H + Xe \rightarrow H_2 + Xe, \quad (11)$$

which has a rate constant $k_{rec} \sim 6 \times 10^{-33}$ cm$^6$/s at $T = 273$ K [26], can be completely neglected, and recombination at the surface of the quartz reactor tube becomes the primary channel of hydrogen atoms loss.

For a low-catalytic quartz discharge tube surface (with an H atom surface loss probability of less than 2% [27]), the diffusion rate to the surface in the steady-state regime is:

$$\frac{1}{[H]}\frac{d[H]}{dt} < -D_{H-Xe}/R^2 \quad (12)$$

The diffusion coefficient of atomic hydrogen in xenon can be estimated as $P \times D_{H-Xe} \approx 0.27$ m$^2$Pa/s [28]. At the xenon pressure $P = 10$ Pa used in this study and with the inner tube radius $R = 2.5$ cm, the diffusion loss rate of atomic hydrogen (12) is at least an order of magnitude lower than its formation rate in processes (9)–(10) (see also the analysis below).

*Electron kinetics in xenon-hydrogen discharge*

The reduced electric field $E/n$ controls the electron drift velocity, $v_e$, in the plasma, the ionization rate, and the rate of excitation of the internal degrees of freedom of molecules by electron impact. Therefore, the $E/n$ parameter is crucial for describing the plasma. Figure 5 shows the dynamics of the reduced field during discharge development. These values were calculated under the assumption of a uniform potential distribution between the electrodes, where $E = U/L$. This approximation is sufficiently accurate for a long discharge tube and large electrode surfaces, which ensure a relatively small potential drop in the cathode and anode layers. The positive column of the discharge was uniform in all cases, as confirmed by the gas excitation intensity along the discharge tube (Figure 2).

Using an estimate of the *E/n* value in the discharge, we can calculate the electron energy distribution function (EEDF), the mean electron energy, and the electron drift velocity (Figure 6). We utilized the BOLSIG+ software to calculate the electron ensemble parameters using the two-term approximation of the Boltzmann equation [29]. A set of electron impact cross-sections for both Xe and $H_2$ from the Siglo database was used [30]. The presence of a tiny amount of molecular gas (hydrogen) in the mixture has a minimal effect on the EEDF and electron swarm parameters. Moreover, as mentioned above, direct electron impact is not the main channel for hydrogen dissociation in this mixture; processes (9)–(10) are much faster and primarily control the rate of hydrogen dissociation in the plasma. Nonetheless, the complete set of excitation, dissociation, and ionization processes by electron impact was included in the calculations for both xenon and molecular hydrogen.

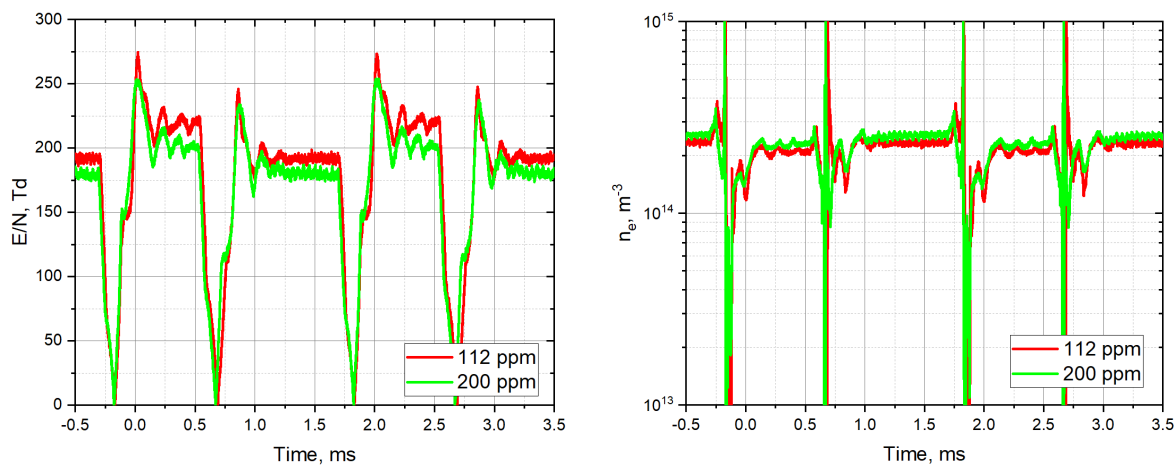

Figure 5. Left: Average reduced electric field in plasma. Right: Electron density in the discharge. Xe+$H_2$, *P* = 10 Pa.

In the range of *E/n* values observed in the experiment (*E/n* = 200±20 Td, Figure 5), the electron energy distribution function and the electron mobility change only slightly. This stability allows for accurate estimation of electron concentration in the discharge using the drift–diffusion approximation. Note that in this range of reduced electric fields, the electron mobility in the xenon – hydrogen mixture decreases slightly with increasing *E/n* (Figure 6). The calculated values of electron drift velocity enable estimation of electron concentration in the discharge using a one-dimensional approximation.

In this 1D approximation, we assume that the concentrations of electrons and H atoms are uniform across the discharge tube cross-section and do not vary with radius, and that the voltage drop in the cathode layer is small compared to that in the positive column of the discharge. Under these assumptions, the total discharge current can be expressed as:

$$I \sim n_e S q_e v_e = n_e S q_e \mu_e E \qquad (13)$$

and

$$n_e \sim I/[S q_e \mu_e E] \qquad (14)$$

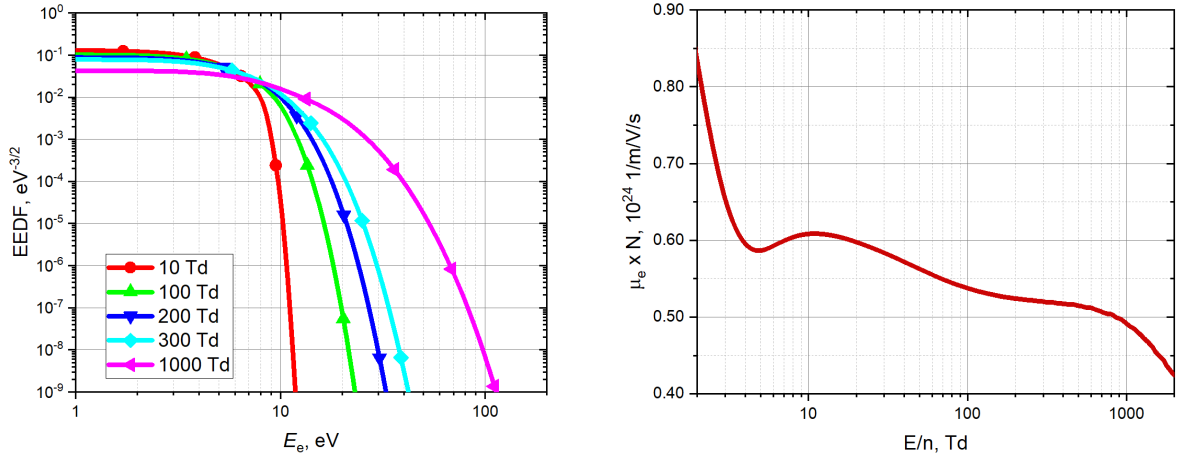

Figure 6. EEDF and electron mobility at different values of the reduced electric field. Mixture Xe + 112ppm $H_2$.

The *E/n* value and the discharge current are known from experimental measurements (Figures 3 and 5), while the electron mobility is obtained from calculations (Figure 6). This allows for an easy estimation of electron concentration in the discharge using equations (13) – (14). The results of this estimation are presented in Figure 5. Within the framework of the considered approximation, estimates show that the concentration of electrons in a discharge in a xenon-hydrogen mixture at $P$ = 10 Pa and a total discharge current of about $I \sim$ 8 mA is $n_e \sim$ 2.3-2.6×$10^{14}$ m$^{-3}$, and the degree of gas ionization is $[n_e]/[n] \sim 10^{-7}$, which is a characteristic for DC glow discharges in atomic gases.

Figure 7 shows the rates of excitation and ionization of gas by electron impact. Note that an increase in the reduced electric field value leads to a rapid increase in the excitation and ionization rate coefficients. Within the range of reduced electric fields corresponding to steady-state phase of the discharge (Figure 5, left), the excitation and ionization rate coefficients can vary by up to twofold. This behavior is due to the increased populating of the high-energy tail of the EEDF with increasing electric field. Fortunately, in a steady-state regime, these variations do not result in significant changes in electron number density or the dissociation degree of molecular hydrogen. The mean electron energies in this range of electric fields range from 5 to 6 eV (Figure 7).

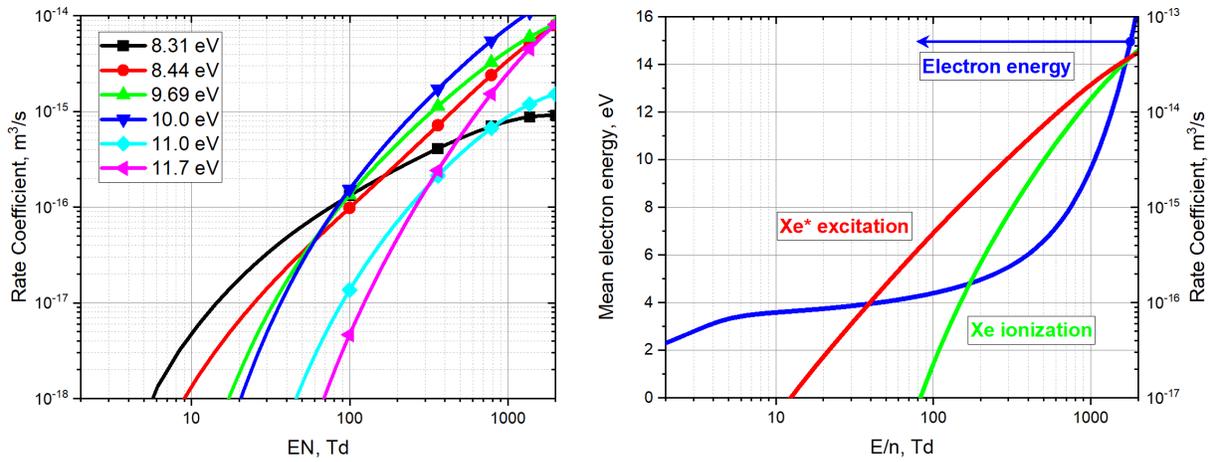

Figure 7. Rate of excitation and ionization of xenon by direct electron impact at different values of the reduced electric field. Mixture Xe + 112 ppm $H_2$.

With knowledge of the electron concentration and the electric field strength in the plasma, it is straightforward to estimate the dissociation rate $R_{diss}$ (reactions (9)-(10)) of molecular hydrogen, the recombination rate $R_{rec}$ (reaction (11)), and the diffusion rate $R_{diff}$ (12) of atomic hydrogen in the plasma (Table 3).

In Table 3, the concentration of atomic hydrogen was calculated assuming complete dissociation of molecular hydrogen. A comparison of the rates of various processes shows that, even in this case, the formation rate of atomic hydrogen far exceeds its loss rate through recombination reactions in the gas and on the discharge chamber walls. Thus, due to this specific combination of conditions, the small admixture of molecular hydrogen in xenon in the discharge is fully dissociated into atomic hydrogen in the ground electronic state. Under these conditions, the concentration of ground-state atomic hydrogen in the plasma is determined solely by the initial concentration of molecular hydrogen in the Xe-$H_2$ mixture. This conclusion is significant for calibration of measurement methods, as it allows for the production of a known concentration of atomic hydrogen that is practically independent of the kinetic processes in the plasma.

Table 3. Mixture composition and rates of major processes in plasma.

| Mixture | [Xe], m$^{-3}$ | [H], m$^{-3}$ | [e], m$^{-3}$ | $R_{diss}$, m$^{-3}$s$^{-1}$ | $R_{rec}$, m$^{-3}$s$^{-1}$ | $R_{diff}$, m$^{-3}$s$^{-1}$ |
|---|---|---|---|---|---|---|
| Xe + 112 ppm $H_2$ | 2.4×10$^{21}$ | 5.4×10$^{17}$ | 2.3×10$^{14}$ | 8.0×10$^{20}$ | 8.0×10$^{12}$ | 2.6×10$^{19}$ |
| Xe + 200 ppm $H_2$ | 2.4×10$^{21}$ | 9.7×10$^{17}$ | 2.6×10$^{14}$ | 9.0×10$^{20}$ | 2.6×10$^{13}$ | 4.7×10$^{19}$ |

*H-fs-TALIF calibration*

For the H-fs-TALIF calibration, the discharge in Xe with an admixture of 112 or 200 ppm of $H_2$ was used (Figure 8). As discussed above, the discharge parameters ensure complete dissociation of hydrogen molecules, resulting in a fixed concentration of H atoms. This enables independent calibration of the parameters in equation (1). These measurements are used to estimate the ratio of the two-photon absorption cross-sections, $\sigma^{(2)}$, for H and Kr.

The TALIF signal generated by neutral H atoms was spatially and temporally resolved using femtosecond two-photon absorption laser-induced fluorescence (fs-TALIF) with a frequency-tunable Ti:sapphire laser (Spectra-Physics Solstice) 90-fs amplified laser system capable of generating up to 200 µJ at 205 nm with a repetition rate of 1 kHz. All experiments were conducted in a low-frequency regime ($f$ = 1 Hz) to ensure the complete mixture replacement between the pulses and the absence of gas heating or perturbations induced by laser radiation. A variable metallic UV silica circular neutral density filter (MKS/Newport 100FS02DV.2) was used to change the pulse energy. To account for the effects of absorption and reflection on the optical elements, the laser pulse energy was measured in front of and behind the focusing lens and the inlet optical window (Figure 8) by ThorLabs Pyroelectric Energy Sensor ES111C (resolution 100 nJ, measurement uncertainty ±5% over a spectral range 190 nm - 25 µm). To reduce the uncertainty, the entire discharge cell 6 and the input window 5 were replaced by a similar window with a laser pulse energy sensor located behind it. Thus, distortions of power measurements due to the presence of the input window of the discharge cell were excluded.

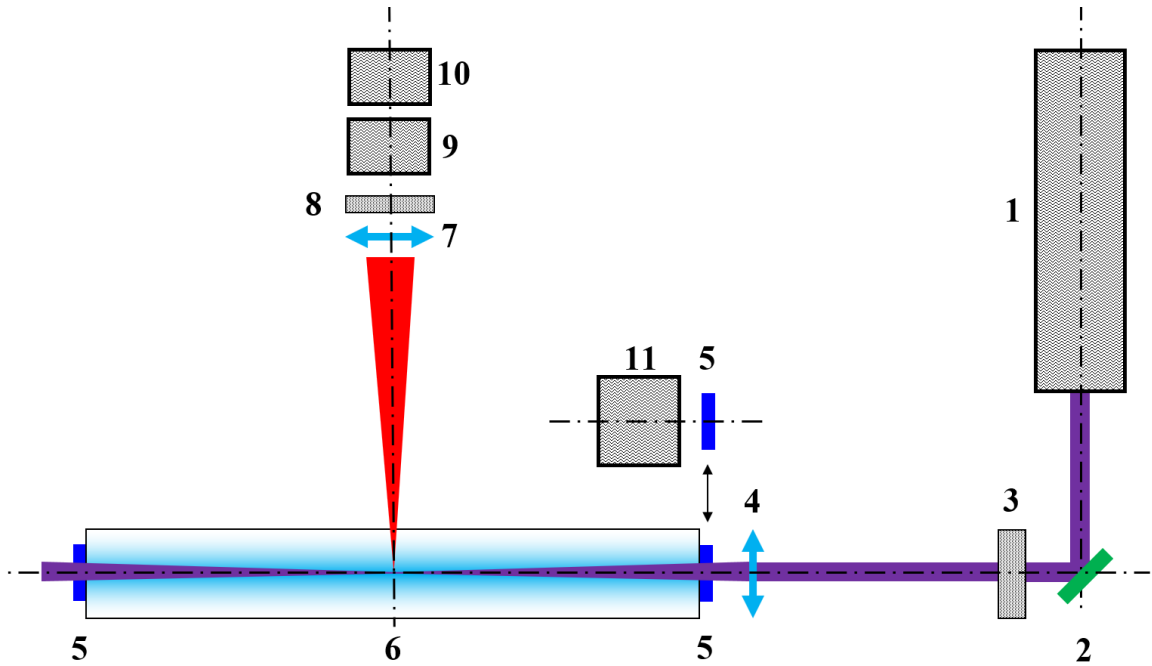

Figure 8. Schematics of the H-fs-TALIF experiment. 1 - 90-fs amplified laser system (200 µJ @ 205 nm), repetition rate $f$ = 1 kHz; 2 – 205 nm dichroic mirror; 3 – calibrated attenuator; 4 - spherical or cylindrical lenses $f$ = 600 mm. 5 – 0.5-mm thick quartz optical windows; 6 – 50 mm diameter, 1-meter-long quartz discharge cell; 7 – interference filter (656 nm or 826 nm) 8 – camera lens; 9 - fast gated image intensifier; 10 - CCD camera; 11 – laser pulse energy sensor.

The laser beam was focused at the center of the discharge cell using spherical or cylindrical lenses with a focal length of $f$ = 600 mm. A back-illuminated CCD camera (Teledyne/Princeton Instruments PIXIS 400B) equipped with a 1.2 ns gated image intensifier (Stanford Computer Optics Quantum Leap N) was used to record the atomic hydrogen fluorescence emission at 656 nm. The standard deviation of the laser pulse energy from pulse to pulse was measured. The pulse-to-pulse deviation was less than 1% @820 nm, less than 2.5% @205 nm and does not affect the results. The measurements of the TALIF signal by ICCD camera were performed by averaging on the CCD 100 successive laser shots (at 1 Hz). This was added on page 18 in the setup section. The images were analyzed by integrating over the same ROI for H and Kr, after subtracting background images in each case.

The image intensifier featured a "High QE RED" photocathode with a maximal sensitivity in the range of 400 – 900 nm. An ultra-narrow bandpass filter (Edmund Optics, 656.3 nm, 1.0 nm FWHM, 90% peak transmission) was used to isolate the emission of hydrogen atoms.

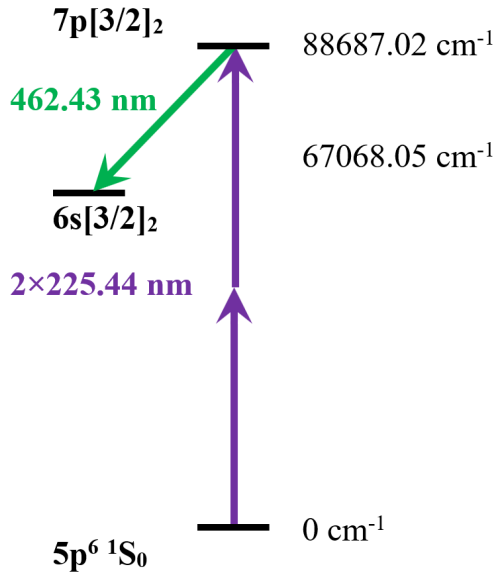

Figure 9. Two-photon excitation energy diagram for TALIF transitions in Xe. The energy levels are given in cm$^{-1}$.

Figure 9 shows the simplified two-photon excitation energy diagram for TALIF transitions in Xe. For example, the scheme does not show the excitation around 224.3 nm and the fluorescence at 834.68, 452.47, and 473.42 nm, which does not change the overall discussion, as the absorption wavelength remains significantly different from the 205 nm used in the present work. The significant difference in the excitation wavelength for the upper Xe levels compared to the two-photon transitions in hydrogen (Figure 1) ensures that there is no excitation of Xe(7p[3/2]$_2$) by the fs-laser pulse and minimal attenuation of laser radiation in xenon due to the low strength of transitions in the 2×205 nm region. Xe atoms can emit in the wavelength range around $\lambda \sim 650$ nm. For example, the transition 9s$^2$[3/2]° → 6p$^2$[5/2] produces radiation at wavelength $\lambda \sim 654.3$ nm, which falls within the sensitivity range of the detection system. However, the radiation lifetime of the 9s$^2$[3/2]° level is very long (on the order of a microsecond) compared to the lifetime of the 3d $^2D_{3/2,5/2}$ level of atomic hydrogen ($\sim 10^{-8}$ s), making it easy to isolate the contribution of these lines to the emission in this wavelength range. Thus, the xenon emission in the 656 nm region could be neglected under the conditions of our experiments. This was confirmed by turning off the discharge in the Xe-H$_2$ mixture, which stopped the atomic hydrogen production and immediately resulted in the disappearance of the TALIF signal. This fact greatly simplifies the interpretation of H-fs-TALIF data in the Xe-H$_2$ mixture.

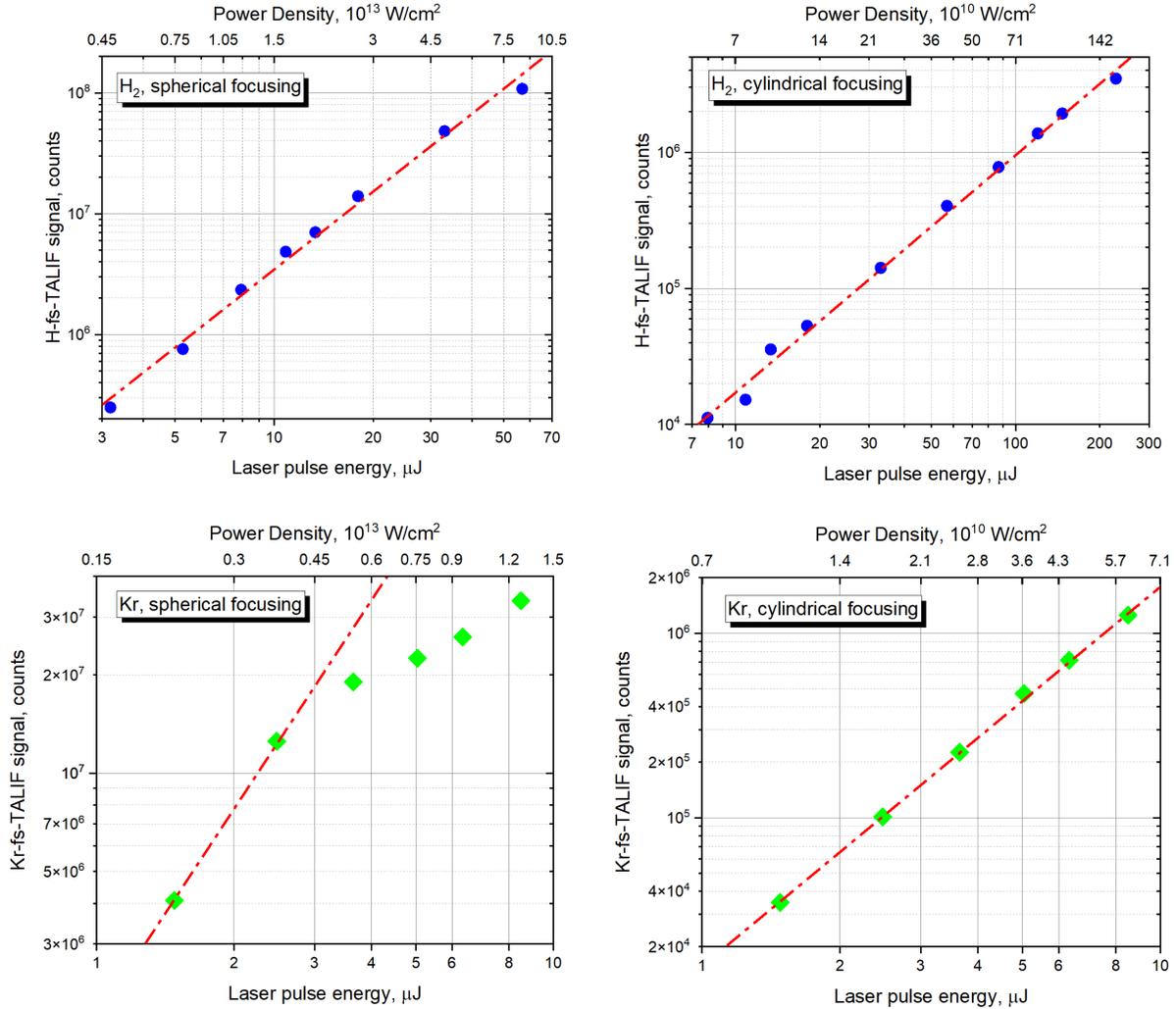

Figure 10. fs-TALIF signals for H (top) and Kr (bottom) as function of pump laser energy for spherical (left) and cylindrical (right) focusing, providing line and sheet TALIF, respectively. Mixture 115 ppm H$_2$ + Xe, total pressure 11.3 Pa (top); pure krypton, total pressure 4 Pa (bottom). Symbols – measurements, solid lines – data approximation; dash-dot lines – quadratic fits through the point (0;0).

The TALIF signal generated by krypton atoms was measured in the same way, using a two-photon absorption wavelength of 204 nm and fluorescence emission at 826 nm. Signal intensities were evaluated from the recorded LIF images using an Edmund Optics interference bandpass filter with a center wavelength of 826 nm, 1.0 nm FWHM and peak transmission of 70%. Similar signals were recorded for both H and Kr, with careful attention to operate in the linear regime, before saturation occurs. The relative quantum efficiency of the image intensifier's photocathode at 654 nm and 826 nm was $QE(654.3)/QE(826.3) = 0.5/0.2$.

The use of a long discharge cell mitigated the issue of two-photon absorption in the windows, as they were located far from the focal point. However, it remains essential to ensure that the TALIF signal is proportional to the density of the species being studied. For accurate absolute density calibration, it is also necessary to prevent signal saturation with the pump laser energy. Since two-photon absorption depends on intensity, the energy saturation threshold is expected to be significantly higher at the lower intensities associated with cylindrical focusing compared to spherical focusing. The TALIF signals were obtained from 2D camera images of the LIF emission. The data was analyzed by integrating over the same-sizes regions of interest for both H and Kr background-subtracted images. Figure 10 shows the fs-TALIF signals for H (top) and Kr (bottom), obtained by varying the laser energy for both spherical (left) and cylindrical (right) focusing. The dash-dot fits in the log-log plots in Figure 10 follow a quadratic dependence on laser pulse energy, characteristic of the two-photon excitation process in TALIF.

Figure 10 demonstrates that focusing with a cylindrical lens effectively avoids saturation due to the much lower intensity of the ~5 mm wide laser sheet compared to the significantly higher intensity achieved with spherical lens focusing (the focal length was identical for all cases $f = 600$ mm). The H-TALIF measurements were conducted with a xenon-hydrogen mixture at $P = 11.3$ Pa, with a hydrogen admixture of 112 or 200 ppm. This means that the partial pressure of atomic hydrogen produced by the discharge was well below 0.1 Pa. For calibration, measurements were performed in 100% Kr gas (cell pressure $P = 4$ Pa, with no discharge). Figure 10 shows that at higher pressures, the TALIF signal obtained from krypton is much stronger and saturates faster.

The quadratic fits indicate that, below the saturation threshold, the fs-TALIF signals exhibit a two-photon excitation dependence on laser field intensity. The radiation intensity depends not only on the laser pulse energy but also on its spatial and temporal distribution, which in turn are influenced by the initial energy distribution within the beam and its focusing. Under the conditions of this study, for hydrogen atoms, the unsaturated regime is below 30 µJ ($I \sim 4.5 \times 10^{13}$ W/cm$^2$) for spherical focusing and not less than 300 µJ ($I \sim 2.1 \times 10^{12}$ W/cm$^2$) for cylindrical focusing; for krypton atoms, it is below 2.5 µJ ($I \sim 3.8 \times 10^{12}$ W/cm$^2$) for spherical focusing and not less than 10 µJ ($I \sim 7 \times 10^{10}$ W/cm$^2$) for cylindrical focusing. Using these fits, we can calculate the coefficients needed for absolute calibration, provided the quenching coefficients are also known. This will be revisited after discussing the lifetime measurements. For all subsequent measurements (lifetime and calibration), cylindrical focusing was used to avoid saturation.

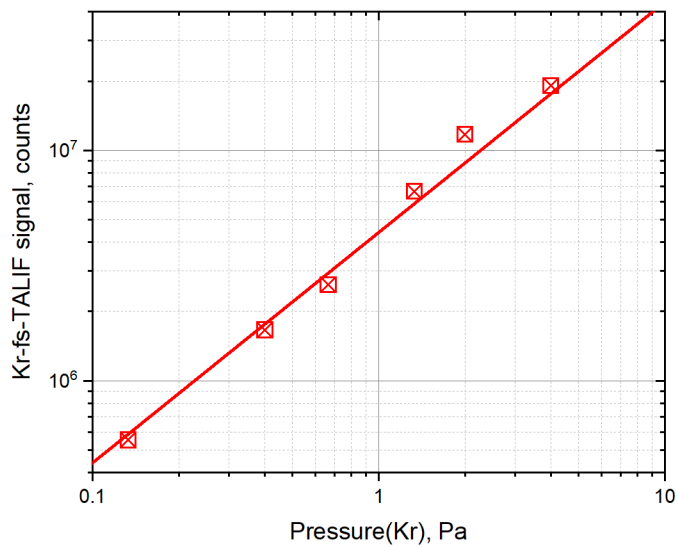

Figure 11. Kr-TALIF signal as function of the Kr pressure. Laser pulse energy 3.65 µJ @ 205 nm. Symbols – measurements, dash-dot line – linear fit.

To calibrate the system accurately for absolute atomic density measurements, we conducted measurements on known densities of Kr, as shown in Figure 11. The Kr-fs-TALIF signal was recorded at a sufficiently low laser energy (3.65 µJ for cylindrical focusing) to ensure an unsaturated response. As shown in Figure 11, the TALIF signal is proportional to atomic density, and the slope from the linear fit is used to calibrate the TALIF signals. For the absolute calibration of H atom density using H-TALIF measurements, it is essential to know the quenching rates, which depend on the gas mixture and the partial pressures of the mixture components. At lower pressures, where collisions can be neglected, atomic emission lifetimes are expected to remain largely unaffected, making these lifetimes straightforward to measure.

The H- and Kr-TALIF data presented above were recorded with a 10 ns integration time. This duration was found to offer the best balance between collecting sufficient luminescence (where a longer integration time is beneficial) and minimizing sensitivity to optical background noise (where a shorter time is preferable). For lifetime measurements, however, we use the shortest available gating time of 1.2 ns in our setup. With ~1 ns resolution, we can measure lifetime by varying the delay between the laser pulse and the ICCD gain gate. Figure 12 presents a series of lifetime measurements performed with H-TALIF (left) and Kr-TALIF (right) at different pressures for the H$_2$-Xe and Kr gases, respectively.

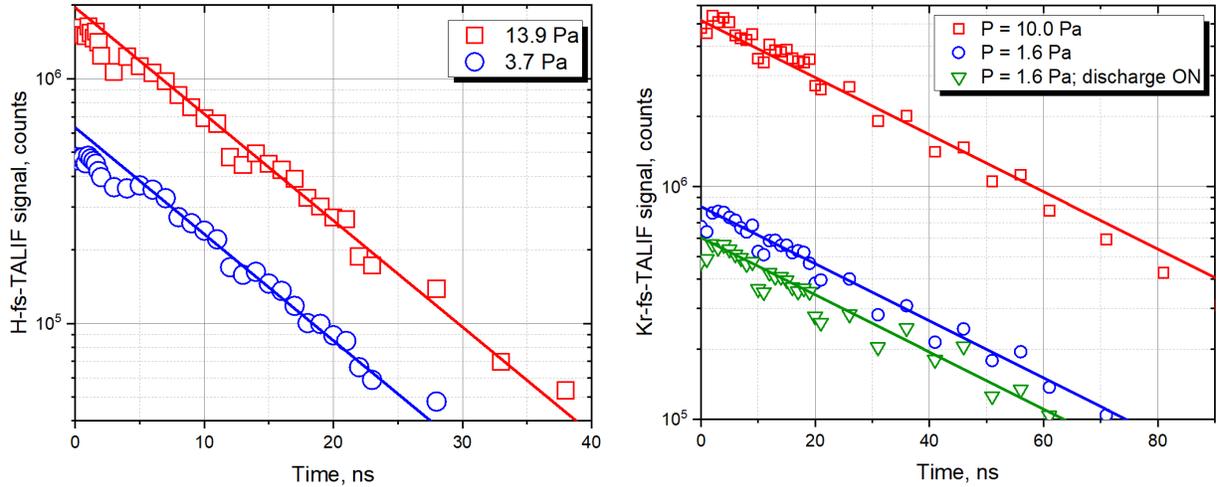

Figure 12. H (left) and Kr (right) emission lifetimes measured with fs-TALIF for several pressures of H$_2$-Xe mixture and Kr, respectively. Symbols – measurements, lines – emission decay fit.

Figure 12 (left) shows that the lifetime of the H emission remains unchanged when increasing pressure from $P = 3.7$ Pa to 13.9 Pa in the 115 ppm molecular hydrogen-xenon mixture, as expected due to the low collision rates at these pressures. The same is observed for Kr, where the measured lifetimes in Figure 12 (right) for Kr at $P = 1.6$ Pa and 10 Pa are very close to the radiative lifetime. We also tested krypton under plasma excitation to measure the effects of ionization (collisional depopulation of excited states by electron impact). As shown in Figure 12 (right), the measured lifetimes align within our experimental error (~0.3 ns), indicating no additional quenching due to gas ionization. This confirms that the discharge itself does not impact our TALIF measurements.

From these measurements, we obtain a lifetime of 10.1±0.3 ns for H, which aligns well with the value reported in reference [11, 14]. This suggests a statistical distribution of the population across the n = 3 sublevels. Notably, measurements conducted at varying total pressures of the mixture – and therefore at different atomic hydrogen concentrations in the discharge – yield consistent values. As the mixture pressure is increased from 3.7 Pa to 13.9 Pa, with complete dissociation of molecular hydrogen in the discharge plasma, the concentration of atomic hydrogen is expected to increase by a factor of 3.7. The observed signal increase was 3.3 times that of the control, differing from the pressure ratio by 10% and falling within the scatter of the experimental data. This concurrence supports the conclusion of complete dissociation of hydrogen in the plasma. A fourfold change in gas pressure in the discharge, along with corresponding changes in both energy density and reduced electric field, would normally lead to significant variations in the degree of gas dissociation – except in the case of complete dissociation, which is largely unaffected by discharge parameters. However, it should be noted that the discharge can alter the overall concentration of gas particles due to temperature changes. This effect is illustrated in Figure 12 (right).

The depopulation rate of the excited state of krypton was measured when excited by a femtosecond laser pulse. Measurements were conducted for krypton pressures of 10 Pa and 1.6 Pa, with the latter divided into two conditions: one without discharge and one with discharge. The data show a consistent depopulation rate for the Kr(5p′[3/2]$_2$) excited state, with a characteristic time of 35.4±1.5 ns. The concurrence of depopulation times suggests that the primary depopulation process is purely radiative, without the involvement of collisional quenching. This conclusion is evident in the case of a pure gas, given the low particle density and collision frequency. In the case of a plasma, however, it is necessary to consider the potential for collisional quenching in superelastic collisions between excited atoms and electrons, as well as the possible depopulation of the Kr(5p′[3/2]$_2$) state through cascade excitation and ionization processes involving high-energy electrons. The identical depopulation times of the excited Kr(5p′[3/2]$_2$) level in both discharge and non-discharge conditions indicate that the role of electron-involved collision processes in this depopulation is minimal.

As observed in the xenon-hydrogen mixture, an increase in krypton pressure is accompanied by a proportional increase in the TALIF signal (see Figures 11 and 12). However, activation of the high-voltage discharge leads to a reduction in the signal, indicating a decrease in gas density within the discharge. This decrease allows for an estimation of the gas temperature increase at the discharge axis under the specified conditions. The observed signal decrease by a factor of 1.3 (Figure 12) suggests an increase in temperature in the discharge by approximately 100 K at a discharge power of $P \approx 7.5$ W.

It should be noted that a comparable analysis of gas temperature with and without discharge in the Xe-H$_2$ mixture is not feasible, as the discharge itself is responsible for generating atomic hydrogen, which is the source of the TALIF signal. However, because the heat flux power is independent of gas density and exhibits only a weak temperature dependence ($P \sim T^{1/2}$), the change in gas temperature and corresponding decrease in gas density at the discharge tube axis can be estimated based solely on the discharge power across different regimes. As stated previously, the discharge power in the Xe-H$_2$ mixture at a pressure of 10 Pa was $P \approx 3.8$ W (Figure 4). Given that the heat transfer coefficient in xenon is 1.7 times lower that of krypton, the temperature increase on the discharge axis in the Xe-H$_2$ mixture can be estimated as $\Delta T = 85$ K, resulting in an approximate 27% decrease in gas density, which should be taken into account in the analysis.

Note that the influence of temperature distribution in the discharge under our experimental conditions was minimal, resulting in only minor corrections to the measured values. However, this is not always the case. For example, a recent study [31] proposed using a high-current pulsed nanosecond discharge to achieve complete dissociation of oxygen in an oxygen-nitrogen mixture. To achieve this effect, the energy density in the discharge reached 1 eV mol$^{-1}$, resulting in an average temperature increase to 1200±400 K across the discharge cross-section, with a nearly parabolic energy distribution profile. Due to significant radial inhomogeneity in energy distribution, strong gas-dynamic perturbations developed during measurement times. These perturbations propagated from the axis to the walls of the discharge tube and back as radial compression waves, changing the gas density about 3 times compared to the reference gas. Consequently, neglecting gas dynamics under high pulse energy density conditions led to incorrect conclusions in [31], significantly overestimating (2-3 times) the cross-section ratio $\sigma^{(2)}$(Xe)/$\sigma^{(2)}$(O).

We estimated the collisional quenching rates $k_q$ ($Q_2 = k_q \times N$, equation (2)) and the radiative lifetime $\tau$ ($A_2 = 1/\tau$) for both H and Kr atoms (Table 4). Note that in the pressure range ($P = 10$ Pa) used in the present work, collisional quenching is almost an order of magnitude slower than radiation depopulation of the upper levels and does not impact the concentration dynamics of the excited states. We also estimated the total optical transmission of the detection system $T$ (interference filter transmittance at the emission wavelength) and the quantum efficiency of the ICCD photocathode $\eta$ in these spectral intervals.

Thus, we have a complete data set for calculating the ratio of the two-photon absorption cross-sections $\sigma^{(2)}$ for H and Kr, using known concentrations of atomic hydrogen, xenon, and krypton in the discharge cell (Table 3). The results of these calculations are presented in Table 5. As shown in Table 5, the ratio of the two-photon absorption cross-sections $\sigma^{(2)}$ for H and Kr when excited by a broadband femtosecond laser differs significantly – by more than a factor of twenty – from data obtained in [5,11] using a narrowband nanosecond lasers.

Table 4. Parameters for H-fs-TALIF calibration.

|  | H | Ref | Kr | Ref |
|---|---|---|---|---|
| $\lambda$ emission, nm | 656.27 | [19] | 826.30 | [19] |
| $\lambda$ laser, nm | 205.08 | [19] | 204.13 | [19] |
| $k_q$, $10^{-10}$ cm$^3$s$^{-1}$ | k(Xe) = 19.8 | [5] | k(Kr) = 1.46 | [5,32] |
|  |  |  | k(Kr) = 1.31±0.065 | [17] |
|  | k(Xe) = 19.8 | Used in this work | k(Kr) = 1.46 | Used in this work |
| $\tau$, ns | 17.6 | [5,32] | 34.1 | [5,32] |
|  | 16.7±0.7 | [17] | 33.6±1.1 | [17] |
|  | 10.0±0.5 | [11,12] | 35.4±2.7 | [11,12] |
|  | 15.7±1.5 | [33] | 26.9 | [12] |
|  | 20.9±0.8 | [21] |  |  |
|  | 13 | [14] | 29 | [14] |
|  | 10.1±0.3 | This work | 35.4±1.5 | This work |
| $A_{2\rightarrow 3}$, s$^{-1}$ | 4.4×10$^7$ | [11] | 2.7×10$^7$ | [11] |
| $T$ | 0.90 | This work | 0.70 | This work |
| $\eta$ | 0.50 | This work | 0.20 | This work |

The accuracy of this estimate relies on the assumption of complete dissociation of hydrogen in the Xe-H$_2$ plasma and the absence of a significant number of hydrogen ions and excited atoms. The kinetic process estimates in this system support the conclusion that the value $\sigma^{(2)}(Kr)/\sigma^{(2)}(H) \sim 0.027$ can be used as calibration data for the fs-system with an accuracy of ±20%.

The large difference in the ratio of two-photon absorption cross-sections $\sigma^{(2)}$ for H and Kr obtained using nanosecond and femtosecond laser excitation is undoubtedly due to the significantly different excitation dynamics of atoms for varying pulse durations in short-pulsed and long-pulsed laser systems, as well as the markedly different spectral energy distributions of these systems. Thus, in [11] a laser system based on a pulsed injection-seeded Nd:YAG laser

Spectra-Physics/Quanta-Ray GCR-230 and a tunable dye laser Spectra-Physics/Quanta-Ray PDL-3 was used to generate the tunable UV radiation. In [5], two different narrow line nanosecond dye lasers were used (Spectron SL400 for 205 nm and Radiant Dyes LDL205 for 225 nm). In our work, we used a frequency-tunable Ti:sapphire laser (Spectra-Physics Solstice) operating at a repetition rate of 1 kHz. The system generates 90 fs pulses with the center wavelength at 820 nm. The measured bandwidth of the fourth harmonic at 205 nm was around 1 nm, corresponding to about 200 cm$^{-1}$. Given the much larger bandwidth of the femtosecond laser, the excitation can be considered uniform over the spectral distribution of the transitions in the atomic system, regardless of pressure or temperature broadening.

Table 5. The ratio of two-photon absorption cross-sections $\sigma^{(2)}$ for H and Kr.

| Method of [H] control | $\dfrac{\sigma^{(2)}(Kr)}{\sigma^{(2)}(H)}$ | Excitation | Ref |
|---|---|---|---|
| Titration with $NO_2$ | $0.62 \pm 50\%$ | ns tunable dye laser, narrow line | [5] |
| Titration with $NO_2$ | 0.60 | ns tunable dye laser, narrow line | [11] |
| DC discharge in Xe + 112 ppm $H_2$ mixture | $0.029 \pm 20\%$ | 90 fs amplified laser system, wide line | This work |
| DC discharge in Xe + 112 ppm $H_2$ mixture | $0.026 \pm 20\%$ | 90 fs amplified laser system, wide line | This work |
| DC discharge in Xe + 200 ppm $H_2$ mixture | $0.025 \pm 20\%$ | 90 fs amplified laser system, wide line | This work |

The substantial differences in excitation dynamics between nanosecond and femtosecond systems (steady-state vs. dynamic regimes) also change the excitation cross sections. Thus, the observed difference in the ratio $\sigma^{(2)}(Kr)/\sigma^{(2)}(H)$ for femtosecond and nanosecond laser systems can be attributed to differences in pulse duration and spectral energy distribution. Note also that, for pairs such as O-Xe (see e.g. the O-fs-TALIF measurements in [1,34-35]), the transition from nanosecond to femtosecond excitation can lead to similar adjustments in calibration.

**Conclusions**

In conclusion, a variable-pressure high-voltage plasma discharge test cell has been developed to analyze the capabilities of H-fs-TALIF for applications in fusion divertors and other environments. The discharge cell operates within a pressure range from a few mPa to hundreds of Pa, with continuous gas flow through the cell to ensure gas exchange and prevent impurities. The plasma parameters in the discharge cell were estimated from the current-voltage characteristics of the discharge. The electric field value, electron concentration, and rates of gas excitation and dissociation were reconstructed using a 1D drift-diffusion approximation. Detailed kinetics analysis indicates that, in Xe-$H_2$ mixtures with low hydrogen content, complete dissociation of molecular hydrogen in the plasma is achieved without significant ionization or excitation of hydrogen atoms. This feature enables the use of this system as a calibrated hydrogen atom source for various applications.

Based on this H-atom source, a new calibration method for H-fs-TALIF was proposed, and the ratio of two-photon absorption cross-sections $\sigma^{(2)}$ for H and Kr was determined. The estimated ratio of the two-photon absorption cross sections $\sigma^{(2)}(Kr)/\sigma^{(2)}(H) = 0.027 \pm 20\%$ for the broadband femtosecond laser excitation is more than twenty times lower than the values obtained with the narrowband nanosecond lasers. This discrepancy is attributed to the significantly different excitation mechanisms in steady-state (nanosecond) and dynamic (femtosecond) regimes, underscoring the need for independent calibration of TALIF measurements in the femtosecond, picosecond, and nanosecond ranges. It should be emphasized that calibration measurements are essential for any experiments aimed at obtaining absolute values. Significant uncertainties in TALIF calibration exist for all types of lasers, whether in the nanosecond, picosecond, or femtosecond ranges, underscoring the importance of conducting new measurements of the ratio of excitation cross-sections across all conditions.

**Acknowledgements**

This work was partially supported by PPPL Award ROA04736-0004 (Program manager Dr. Ahmed Diallo, Princeton Plasma Physics Laboratory, Princeton), DOE Award DE-AC02-09CH11466, and DOE Award DE-SC0024530 (Program manager Dr. Nirmol).